\documentclass[Physsubmission, Phys]{SciPost}
\makeatletter
\let\c@lofdepth\relax
\let\c@lotdepth\relax
\makeatother
\usepackage{subfigure}

\hypersetup{
    colorlinks,
    linkcolor={red!50!black},
    citecolor={blue!50!black},
    urlcolor={blue!80!black}
}

\usepackage[bitstream-charter]{mathdesign}
\DeclareSymbolFont{usualmathcal}{OMS}{cmsy}{m}{n}
\DeclareSymbolFontAlphabet{\mathcal}{usualmathcal}

\begin{document}

\begin{center}{\Large \textbf{
Multi-differential studies to explore strangeness enhancement in pp with ALICE at the LHC}}\end{center}

\begin{center}
Francesca Ercolessi\textsuperscript{1} on behalf of the ALICE Collaboration
\end{center}

\begin{center}
{\bf 1} University and INFN, Bologna, Italy.
\\
*francesca.ercolessi@cern.ch
\end{center}

\begin{center}
\today
\end{center}



\definecolor{palegray}{gray}{0.95}
\begin{center}
\colorbox{palegray}{
  \begin{tabular}{rr}
  \begin{minipage}{0.1\textwidth}
    \includegraphics[width=30mm]{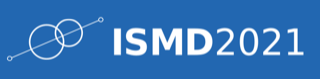}
  \end{minipage}
  &
  \begin{minipage}{0.75\textwidth}
    \begin{center}
    {\it 50th International Symposium on Multiparticle Dynamics}\\ {\it (ISMD2021)}\\
    {\it 12-16 July 2021} \\
    \doi{10.21468/SciPostPhysProc.?}\\
    \end{center}
  \end{minipage}
\end{tabular}
}
\end{center}

\section*{Abstract}
{\bf
The study of energy and multiplicity dependence of strange hadron production in proton-proton collisions provides a powerful tool to understand similarities and differences between small and large collision systems. In order to better understand the origin of strangeness enhancement in pp new multi-differential analyses have been performed. The first separates the contribution of soft and hard processes, such as jets, to strange hadron production through two-particle correlation techniques. The second exploits the concept of the effective energy available for particle production in the event, which is estimated by an anti-correlation with the energy deposited in ALICE's Zero Degree Calorimeters. The results indicate that strangeness production emerges from the growth of the underlying event and confirm it is related to the final state multiplicity.
}

\section{Introduction}
\label{sec:intro}
The primary goal of the ALICE experiment is to study the physics of strongly interacting matter, including the properties of the quark-gluon plasma (QGP), which is expected to be produced in ultra-relativistic heavy-ion collisions. The relative production of strange hadrons with respect to non-strange hadrons in heavy-ion collisions is an important observable to investigate the strongly interacting medium created in the interaction. Strangeness enhancement was one of the first proposed signatures of QGP formation \cite{Rafelski_PRL_1982}, which is expected to lead to an increased production of strange hadrons relative to pp interactions, following a hierarchy related to the strangeness content of the particle. This phenomenon was observed for the first time at SPS \cite{SPS_PLB_1999}, then at RHIC \cite{RHIC_PRL_2012} and later at the LHC \cite{LHC_PLB_2014} at increasing collision energies.
However, the ALICE experiment first observed that small systems, such as p--Pb and pp, show striking similarities with A--A collisions when multiplicity dependent studies are performed.
The ratio of strange particle yields to pion yields as a function of the charged-particle multiplicity produced in the event is found to increase across multiplicities, evolving smoothly within different collision systems and energies \cite{ALICE_NP_2017}.
The results can be compared to various model predictions, from a statistical hadronization description using the canonical suppression approach \cite{CSM_PRC_2019}, to rope hadronization models including colour reconnection effects \cite{CR_PRD_2019}, to two-component (core-corona) models \cite{CC_PRC_2020}. However, the mechanisms responsible for this phenomenon in small systems are still unclear and further experimental studies are needed.
\section{Multi-differential studies}
The ALICE experiment detects $K_{\textrm{S}}^{0}$, $\Lambda$, $\Xi$ and $\Omega$ baryons by reconstructing their weak decay tracks in the central pseudorapidity region. The main subdetectors deployed for particle identification (PID) and tracking include the six-layer high resolution Inner Tracking  System (ITS) and the large volume  Time  Projection  Chamber (TPC). The Time-Of-Flight detector (TOF) is exploited to suppress the out-of-bunch pileup. Two forward detectors placed on both sides of the ALICE interaction point, the V0 and the Zero Degree Calorimeters (ZDC), are used to classify events in multiplicity and energy event classes, respectively. A detailed description of the ALICE apparatus and its performance can be found in \cite{ALICE_JINST_2008}.\\ 
In these proceedings a selection of the ALICE results that explore strangeness production in proton-proton collisions at $\sqrt{s}$ $=$ $13$ TeV will be discussed. 
\subsection{Strangeness in and out of jets}
\begin{figure}[h!]
	\centering
	\includegraphics[width=0.48\paperwidth]{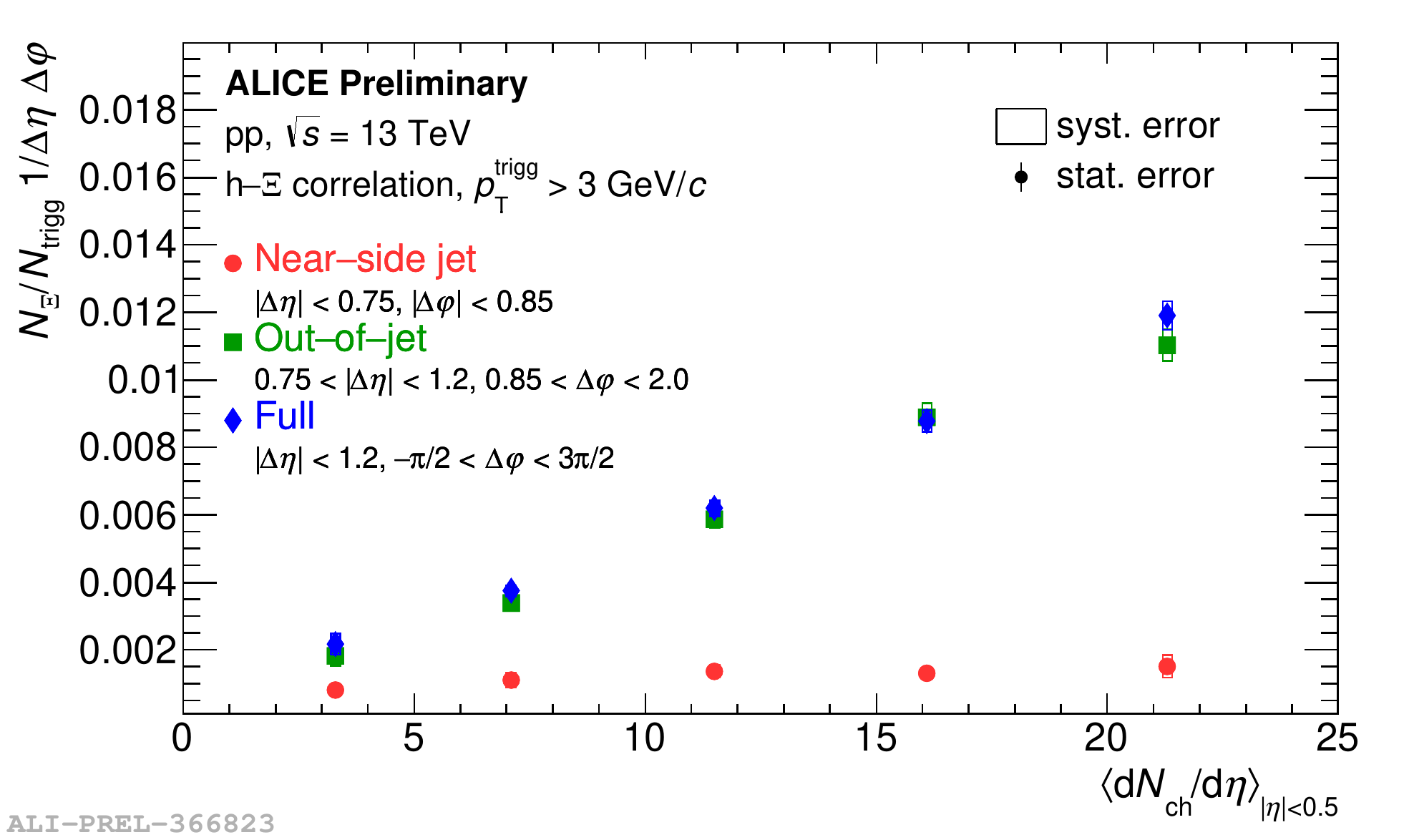}
	\caption{$\Xi$ yields per trigger particle and per unit of $\Delta\eta\Delta\phi$ as a function of the charged-particle multiplicity produced at midrapidity.}
	\label{YieldsvsMultK0s}       
\end{figure}

\begin{figure}[h!]
	\centering
	\includegraphics[width=0.48\paperwidth]{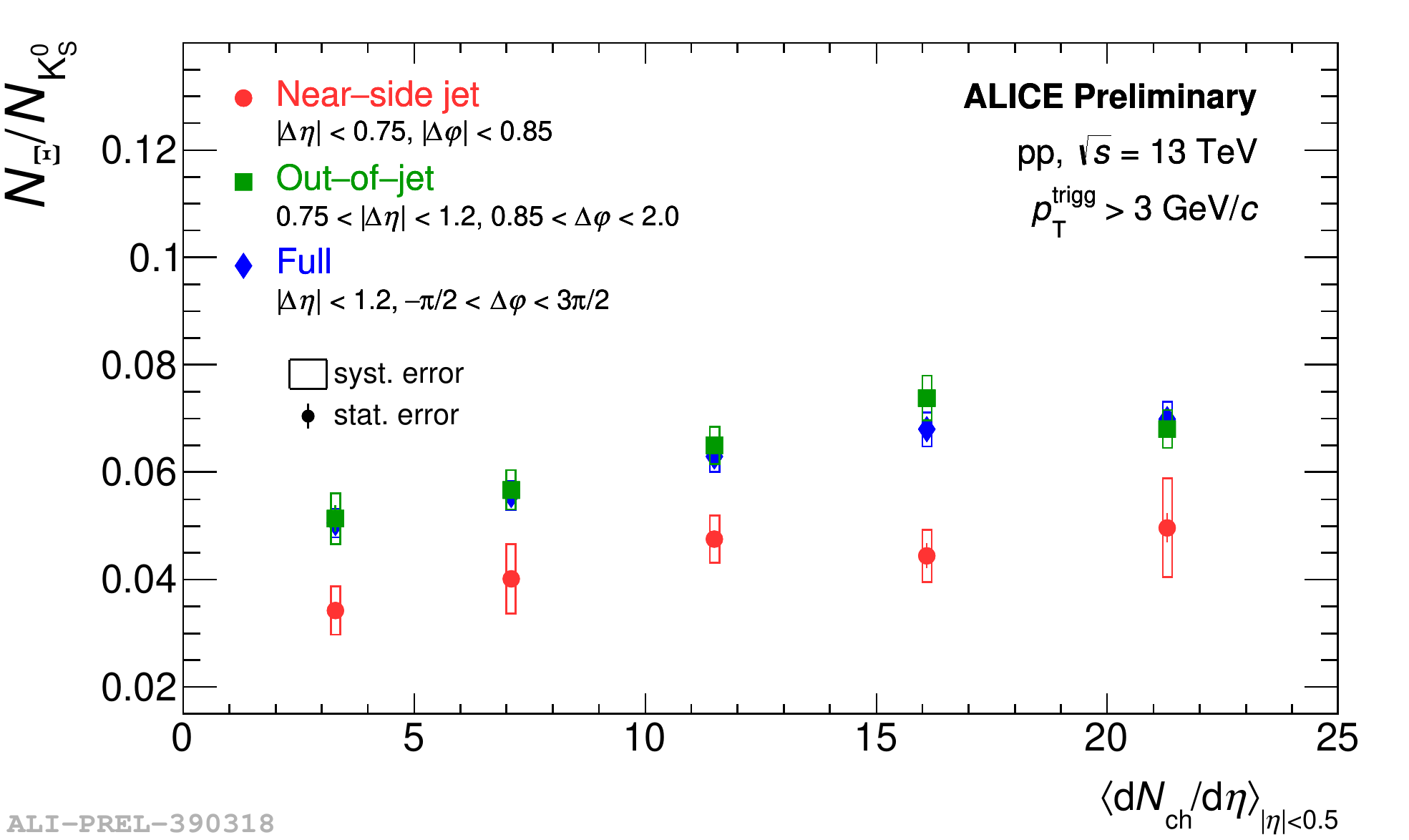}
	\caption{Ratio between $\Xi$ and $\textrm{K}_{\textrm{S}}^{0}$ yields as a function of charged particle multiplicity at midrapidity.}
	\label{RatioXiK0S}       
\end{figure}

The processes involved in particle production in pp collisions can be classified into hard and soft ones, depending on the momentum transfer. The relative contribution of hard and soft processes to strangeness production in pp is still not clear and can be studied through techniques involving full-jet reconstruction or two-particle correlations.\\
In this analysis, ALICE exploits the angular correlation method to separate  $\textrm{K}_{\textrm{S}}^{0}$ and $\Xi$ hadrons produced in jets (hard processes) from those produced out-of-jets (soft processes). This technique is based on the fact that particles produced in the near-side jet region are characterised by a small angular separation from the leading particle of the jet, which is identified as the particle with the highest transverse momentum in the collision and with $p_{\textrm{T}}$ $>$ $3$ GeV/$c$. The angular correlation distribution of the jet trigger particle and the associated strange hadron is studied in the $(\Delta\eta,\Delta\phi)$ plane, which is divided into a near-side-jet region, an out-of-jet region, and a full inclusive region. The near-side-jet contribution is obtained subtracting the full and out-of-jet in the corresponding $(\Delta\eta,\Delta\phi)$ region. The $\Xi$ yields per trigger particle and per unit of $\Delta\eta\Delta\phi$ are displayed in Fig. \ref{YieldsvsMultK0s} as a function of the charged-particle multiplicity produced at midrapidity.
The full and out-of-jet yields increase with multiplicity, while the near-side-jet yields show a very mild to no dependence on particle production at midrapidity. Similar results are obtained for $\textrm{K}_{\textrm{S}}^{0}$ yields.
\\The ratio between $\Xi$ and $\textrm{K}_{\textrm{S}}^{0}$ yields as a function of charged particle production at midrapidity is shown in Fig. \ref{RatioXiK0S}. The out-of-jet and the full yield ratios increase with multiplicity, confirming the observation that strangeness enhancement is larger for particles with larger strangeness content. The near-side-jet yield ratio shows a hint of increase with multiplicity as well. These results suggest that out-of-jet (soft) processes are the dominant contribution to strange particle production in pp collisions.

\subsection{Influence of initial and final state effects on strangeness production}
\begin{figure}[h!]
	\centering
	\subfigure[]{\includegraphics[width=0.32\paperwidth]{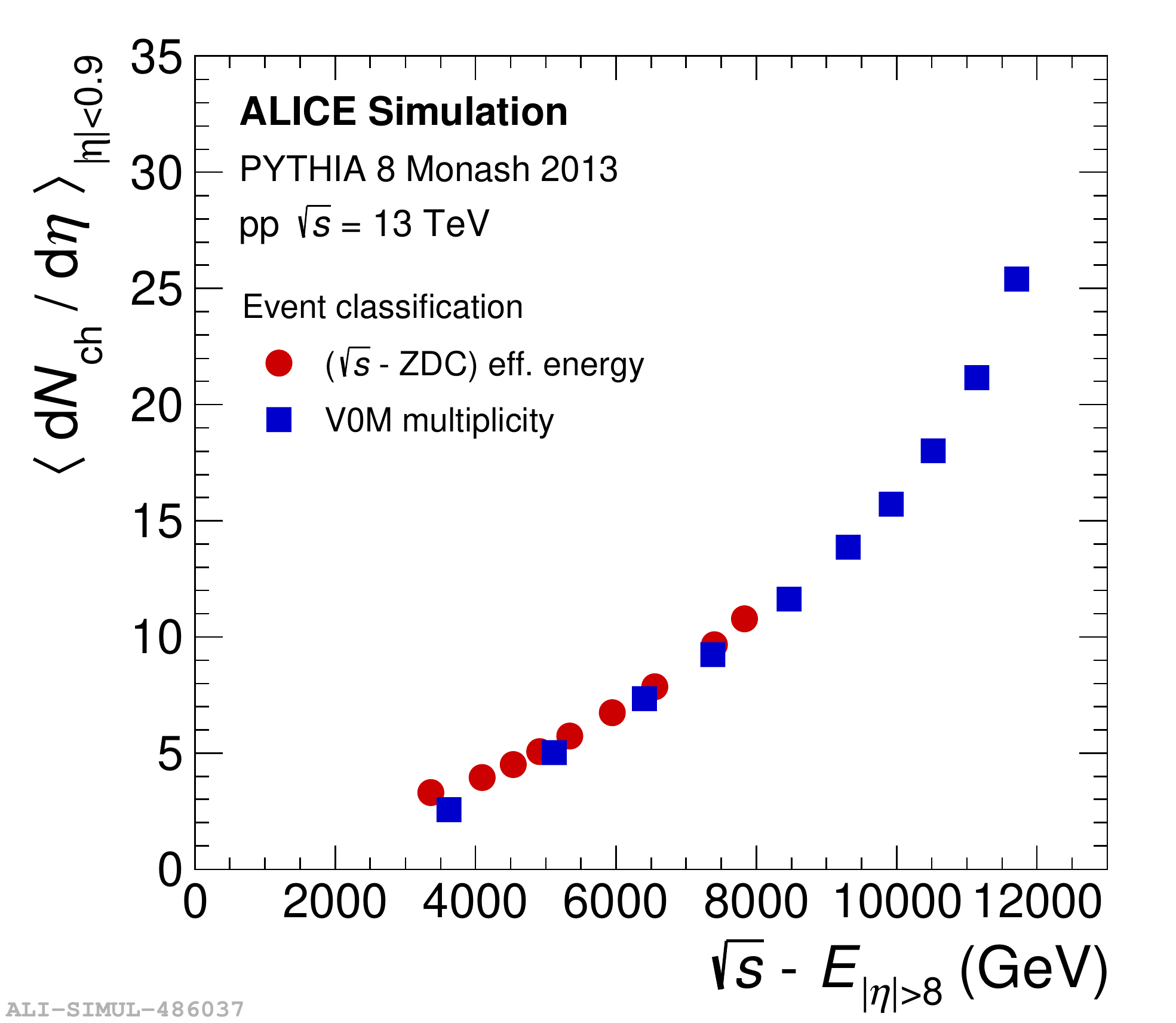}}
	\subfigure[]{\includegraphics[width=0.37\paperwidth]{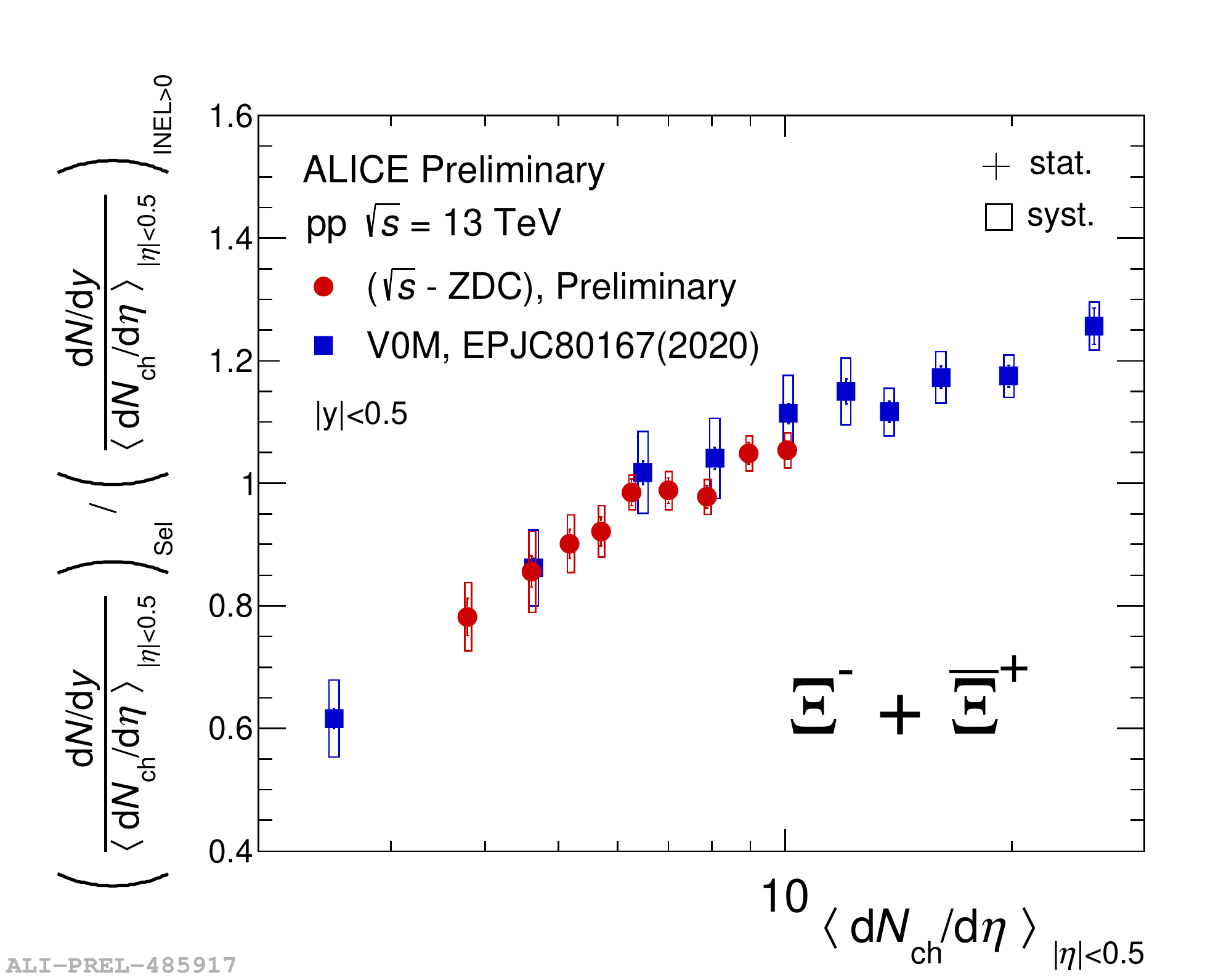}}
	\caption{(a) Multiplicity and effective energy correlation in PYTHIA 8 using V0 and ZDC event classes. (b) Ratio of $\Xi$ yields to the charged-particle multiplicity (self-normalized to INEL>0) as a function of particle production at midrapidity, using V0 and ZDC standalone selections.}
	\label{XiPythia8Standalone}       
\end{figure}

\begin{figure}[h!]
	\centering
	\includegraphics[width=0.38\paperwidth]{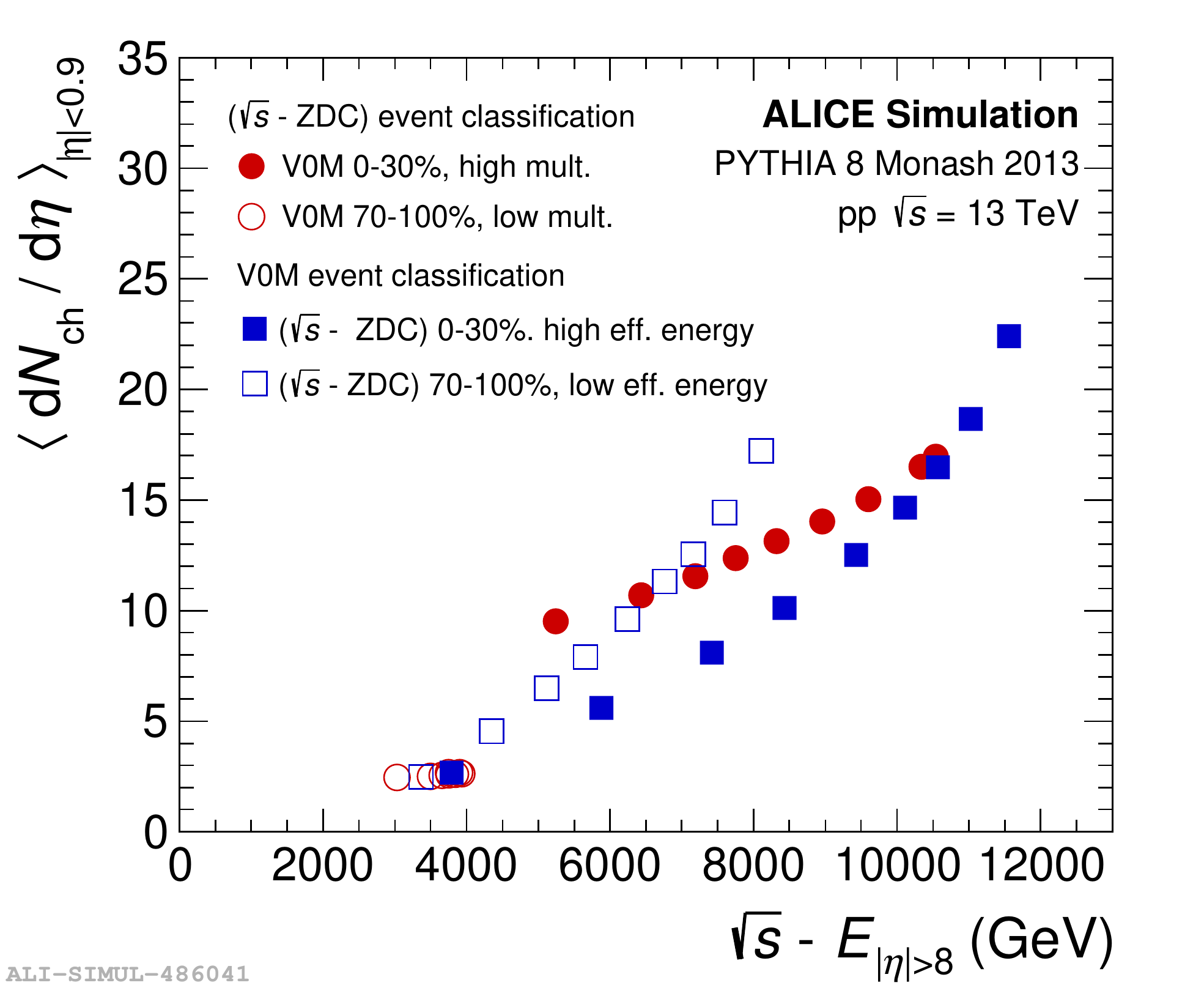}
	\caption{Multiplicity and effective energy correlation in PYTHIA 8 using V0-ZDC combined classes. }
	\label{Pythia8MultiDiff}       
\end{figure}

\begin{figure}[h!]
	\centering
	\subfigure[]{\includegraphics[width=0.35\paperwidth]{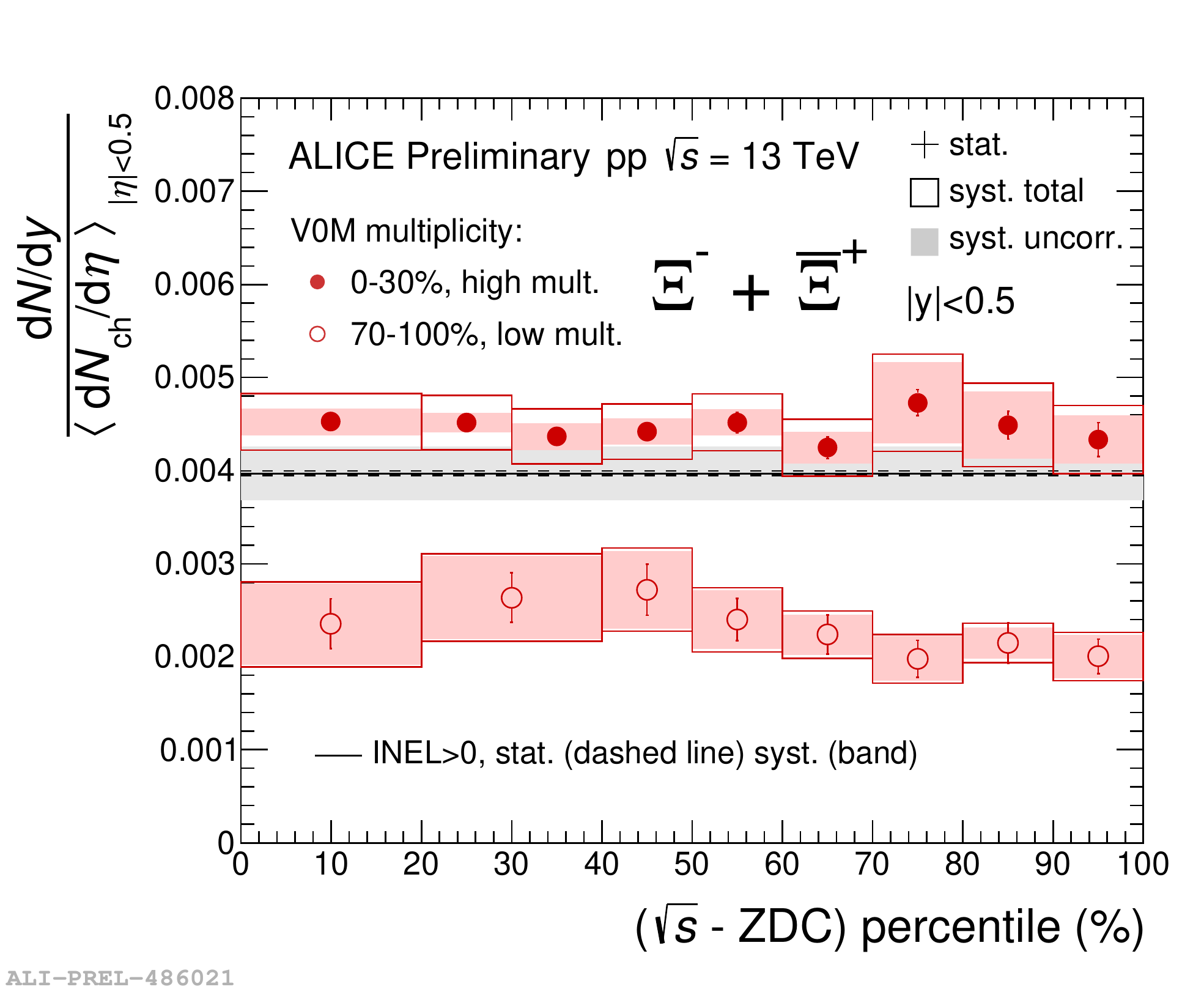}}
	\subfigure[]{\includegraphics[width=0.355\paperwidth]{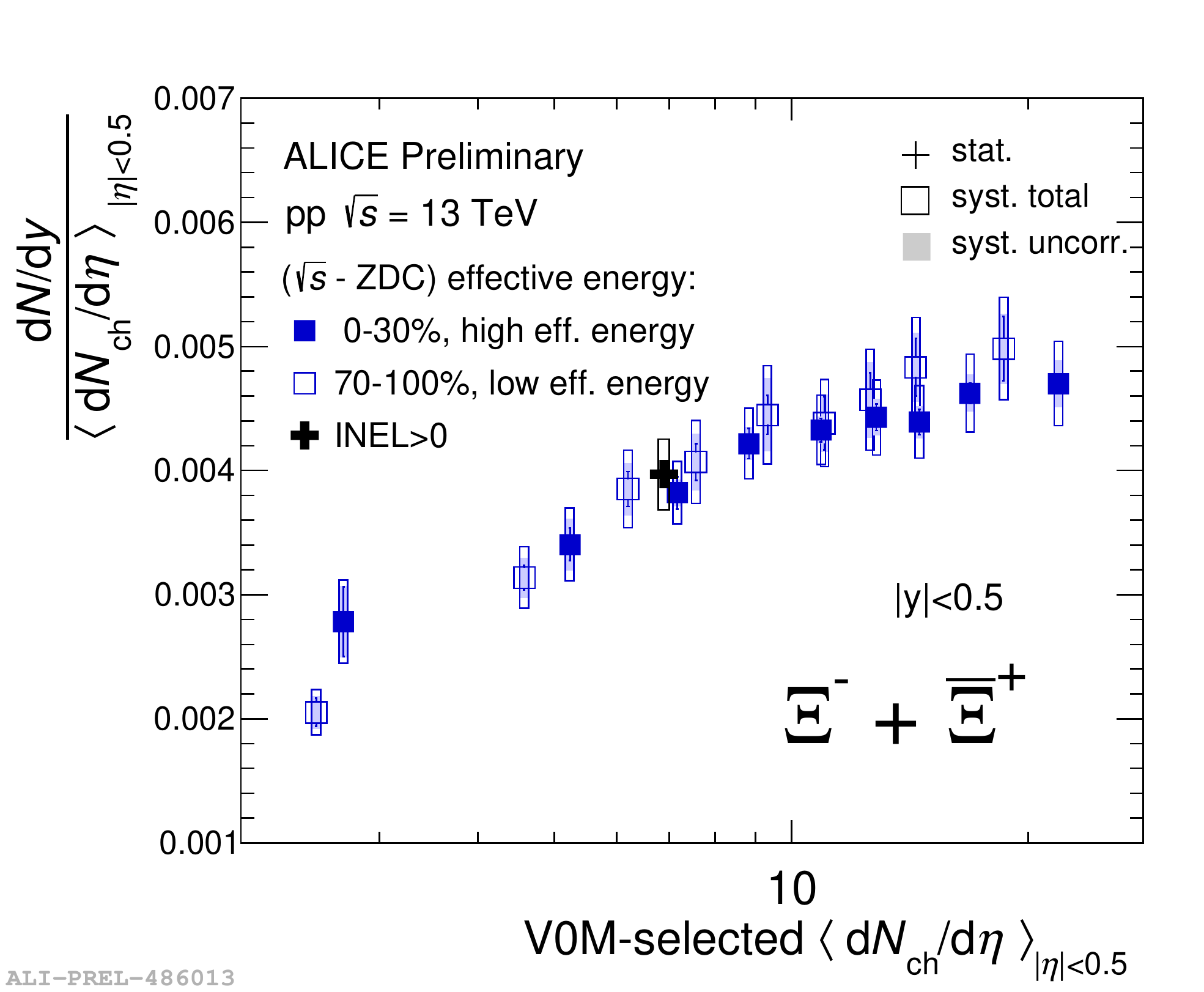}}
	\caption{Ratio of $\Xi$ yields to the charged-particle multiplicity extracted using V0-ZDC combined selections. (a) Events are selected using ZDC event classes, fixing the multiplicity through the V0 estimator. (b) Events are selected using V0 event classes, fixing the effective energy through the ZDC estimator.}
	\label{XiMultiDiff}       
\end{figure}

The multiplicity distribution of charged particles is characteristic of the hadronic final state of any collision, but it also reflects its initial dynamics being strongly correlated with the energy effectively available for particle production in the initial stages of the interaction (effective energy). In pp collisions, the effective energy is reduced with respect to the full centre-of-mass energy due to the leading baryon effect, which consists of a high probability to emit baryons at forward rapidity, namely with large longitudinal momenta along the direction of the incident beams. ALICE's Zero Degree Calorimeters are able to detect the energy deposits of the forward leading baryons in each event hemisphere as $E_{\textrm{effective}} = \sqrt{s} - E_{\textrm{leading}}$. More details can be found in \cite{EE_EPJ_2007} and references therein. In this analysis, the ZDC and V0 detectors are used to classify events in effective energy and multiplicity percentile classes, respectively. The goal of this study is to disentangle initial and final state effects on strange hadron production. Multiplicity and effective energy are two correlated quantities, in particular recent ALICE results showed that the forward energy deposit in the ZDC is anti-correlated with particle production at midrapidity \cite{ALICE_arxiv_2021}. This correlation is studied in this analysis using the PYTHIA 8 event generator and displayed in Fig. \ref{XiPythia8Standalone} (a), which shows that events selected using a V0 or ZDC\footnote{
	The ZDC based percentile estimator is labelled as $\sqrt{s} - ZDC$ and it represents the percentile classes of the quantity $\sqrt{s} - E_{\textrm{ZDC}}$.} based estimator alone are sensitive to both initial and final state.
Fig. \ref{XiPythia8Standalone} (b) displays the ratio of $\Xi$ yields to the charged-particle multiplicity in the event (self-normalized to INEL>0\footnote{INEL>0 is a conventional event class which contains events with at least one charged particle in |$\eta$|$<$ 1.}) selected using V0 or ZDC, as a function of particle production at midrapidity. The $\Xi$ production is found to increase with multiplicity following a similar trend, independent of the estimator used. 
\\In order to separate initial and final state contributions to strange particle production, V0-ZDC combined classes are exploited. Their discriminating power in terms of effective energy and multiplicity is studied in Fig. \ref{Pythia8MultiDiff} via the PYTHIA 8 event generator. Events are selected using ZDC percentile selections, by fixing the multiplicity through the V0 estimator in a high and a low range (red points), and vice-versa using V0 percentile selections, by fixing the effective energy through the ZDC estimator to a high and a low range (blue points).  The ratio of $\Xi$ yields to the charged-particle multiplicity selected using V0-ZDC combined event classes is shown in Fig. \ref{XiMultiDiff} (a) and (b). When fixing the multiplicity through the V0 estimator and selecting events in ZDC percentile classes (Fig. \ref{XiMultiDiff} (a)) the points are distributed in a flat trend showing no dependence on the energy percentile estimator. The different V0 selection is responsible for the enhancement observed between the two sets of points. Fig. \ref{XiMultiDiff} (b) shows the ratio of $\Xi$ yields to the particle multiplicity selected using V0 event classes, fixing the effective energy through the ZDC estimator in a high and a low range. The two trends are consistent within the experimental uncertainties and the selection in effective energy does not influence the observed enhancement with multiplicity. These preliminary results suggest that the effective energy does not play a significant role in the strangeness enhancement observed in pp collisions, confirming a strong role of the final state multiplicity. More studies are ongoing in order to improve the discriminatory power of multi-differential selections in terms of multiplicity and effective energy.  

\section{Conclusion}
The ALICE Collaboration has provided a comprehensive set of results on the production of strange hadrons in pp collisions. These results show striking commonalities with heavy-ion collisions, although the mechanisms responsible of the strangeness enhancement with multiplicity in small systems are still not fully understood. 
The set of preliminary results presented in these proceedings exploit multi-differential approaches to study strange hadron production in pp collisions at $\sqrt{s}$ = 13 TeV. The first analysis separates the contribution of soft and hard processes, such as jets, to strangeness production through two-particle correlation techniques. Soft out-of-jet processes are found to be the dominant contribution, with respect to in-jet hard processes. The second analysis aims at disentangling initial and final state effects on strange hadron production selecting the events in multiplicity and effective energy classes. The results confirm that strangeness enhancement is connected to the final particle multiplicity produced in the collision and suggest no significant dependence on the initial effective energy.

\newpage


\nolinenumbers

\end{document}